\documentclass{physeauth}
\usepackage{graphicx}
\usepackage{amsmath}
\usepackage{amssymb}

\def\kF{k_{\text{F}}}

\def\NF{N_{\text{F}}}
\def\me{m_{\text{e}}}
\def\ne{n_{\text{e}}}

\def\sgn{{\text{sgn\,}}}
\def\be{\begin{equation}}
\def\ee{\end{equation}}
\def\bea{\begin{eqnarray}}
\def\eea{\end{eqnarray}}
\def\bse{\begin{subequations}}
\def\ese{\end{subequations}}
\def\b{\bibitem}

\begin{document}

\begin{frontmatter}

\title{Electronic Transport at Low Temperatures: Diagrammatic Approach}

\author[or]{D. Belitz \thanksref{thank1}},
\author[md]{T.R. Kirkpatrick}

\address[or]{Department of Physics and Institute of Theoretical Science,
             University of Oregon, Eugene, OR 97403, USA}
\address[md]{Institute for Physical Science and Technology and Department of
             Physics, University of Maryland, College Park, MD 20742, USA}

\thanks[thank1]{Corresponding author. E-mail: dbelitz@uoregon.edu}

\begin{abstract}
  We prove that a diagrammatic evaluation of the Kubo formula for the electronic
  transport conductivity due the exchange of bosonic excitations, in the usual
  conserving ladder approximation, yields a result consistent with the Boltzmann
  equation. In particular, we show that an uncontrolled approximation that has
  been used to solve the integral equation for the vertex function is unnecessary.
  An exact solution of the integral equation yields the same asymptotic
  low-temperature behavior as the approximate one, albeit with a different
  prefactor, and agrees with the temperature dependence of the Boltzmann
  solution. Examples considered are electron scattering from acoustic phonons, and
  from helimagnons in helimagnets.
\end{abstract}

\end{frontmatter}


\section{Introduction}
\label{sec:1}

The earliest systematic treatment of electronic transport in solids was
obtained by using the Boltzmann equation. The latter is known to become exact
in the limit of vanishing scatterer density. This implies that it becomes exact
in the limit of low temperatures for scattering processes that involve
excitations, such as phonons, that get frozen out in the limit of zero
temperature. For most problems, the Boltzmann equation cannot be solved in
closed form, but systematic schemes of successive approximations exist; for
instance, the method of variational solutions \cite{Wilson_1954__1}. Usually,
the lowest order variational solution yields the exact temperature dependence
of the transport coefficients; one of the best known examples is the
Bloch-Gr{\"u}neisen law for the electrical resistivity $\rho$ due to acoustic
phonons, which states that the leading temperature dependence of $\rho$ is
given by $\rho \propto T^5$ \cite{Wilson_1954__1}.

A different approach to transport phenomena uses the Kubo formula
\cite{Kubo_1957__2}, which expresses the transport coefficients in terms of
equilibrium time correlation functions that can be analyzed by means of
many-body perturbation theory. For the electrical conductivity, the relevant
correlation function is the current-current correlation. It is known that an
analysis that respects the conservation of charge requires a consistent
treatment of self-energy and vertex-correction diagrams. As a result, the
transport relaxation rate is different from the single-particle relaxation
rate, and the former is much harder to calculate than the latter. The problem
can be cast in the form of an integral equation for the energy-dependent
transport relaxation rate $1/\tau_{\text{tr}}(\epsilon)$ that is not easy to
solve \cite{Holstein_1964__2'}. An approximation that ignores the energy
dependence of $1/\tau_{\text{tr}}(\epsilon)$ \cite{Mahan_1981__3} reduces the
integral equation to an algebraic equation and leads to a result that
reproduces the Bloch-Gr{\"u}neisen result for the electrical transport rate due
to electron-phonon scattering. This is in contrast to the single-particle rate
due to phonons, which is proportional to $T^3$. However, because of the known
energy dependence of the electronic self energy, which can be quite pronounced,
this approximation has been questioned, and even the very validity of the $T^5$
law as derived from many-body perturbation theory has been called into
question\cite{Mahan_1981__3}. While such considerations are not consistent with
what is known from the Boltzmann equation, and thus should cast more doubt on
the Kubo formula or its evaluation then on the Bloch-Gr{\"u}neisen law, they
have led to a popular notion that the $T^5$ law is not firmly established. In
any case, reducing the integral equation to an algebraic one is indeed an
uncontrolled approximation that can be justified only by reference to the
Boltzmann equation, not within the diagrammatic formalism itself.

Scattering by excitations other than phonons leads to analogous questions. For
instance, in the ordered phase of helical magnets such as MnSi the Goldstone
modes associated with the helical order -- helimagnons -- lead to a
single-particle relaxation rate proportional to $T^{3/2}$, and the
approximation mentioned above suggests a transport rate proportional to
$T^{5/2}$ \cite{BKR__4}. While this result is extremely plausible by analogy to
the electron-phonon problem, and can be checked by solving the Boltzmann
equation, it would nevertheless be very valuable to ascertain it directly
within the many-body formalism. The purpose of this paper is to provide this
missing link: We prove that the integral equation for the transport relaxation
rate has a solution that is proportional to $T^5$ for the electron-phonon
problem, and to $T^{5/2}$ for the electron-helimagnon problem.

\section{Integral equation for the transport relaxation rate}
\label{sec:2}

Consider electrons scattered by excitations that lead to an effective
scattering potential with a spectrum $V''({\bf p},u)$, where ${\bf p}$ and $u$
denote the wave vector and the frequency, respectively. Standard developments
\cite{Mahan_1981__3} lead to the following expression for the electrical
conductivity $\sigma$:
\begin{equation}
\sigma = \frac{\ne e^2}{2m_{\text{e}}} \int_{-\infty}^{\infty} d\epsilon\
\left(\frac{-\partial f}{\partial\epsilon}\right)\,
\frac{\Lambda(\epsilon)}{\Gamma_0(\epsilon)}\ ,
\label{eq:1}
\end{equation}
with $\ne$ the electron density and $\me$ the electron mass. If one defines
\bea
{\bar V}''_m(u) &=& \frac{1}{2\kF^2} \int_0^{2\kF}
dp\,p\,\left(\frac{p^2}{2\kF^2}\right)^m\ V''({\bf p},u) \nonumber\\
&&\hskip 120pt (m=0,1),
\label{eq:2}
\eea
then the energy-dependent single-particle relaxation rate $\Gamma_0$ is given
by
\be
\Gamma_0(\epsilon) = -\NF \int_{-\infty}^{\infty} du\
\left[n\left(\frac{u}{T}\right) + f\left(\frac{u+\epsilon}{T}\right)\right]\,
{\bar V}''_0(u)\ ,
\label{eq:3}
\ee
with $\NF$ the electronic density of states per spin at the Fermi surface, and
$n(x)$ and $f(x)$ the Bose and Fermi distribution functions, respectively. For
a given ${\bar V}''_0$, $\Gamma_0$ is an explicitly known function of $T$ and
$\epsilon/T$. The vertex part $\Lambda$ is given as the solution of the
integral equation
\bea
\Lambda(\epsilon) &=& 1 - \NF \int_{-\infty}^{\infty} du\
\left[n\left(\frac{u}{T}\right) + f\left(\frac{u+\epsilon}{T}\right)\right]\,
\nonumber\\
&&\hskip 40pt \times \left[{\bar V}''_0(u) - {\bar V}''_1(u)\right]\
\frac{\Lambda(\epsilon+u)}{\Gamma_0(\epsilon+u)}\ .
\label{eq:4}
\eea
The transport relaxation rate $\Gamma_{\text{tr}}(\epsilon) \equiv
\Lambda(\epsilon)/\Gamma_0(\epsilon)$ is the solution of
\be
\frac{1}{\Gamma_{\text{tr}}(\epsilon)} = \frac{1}{\Gamma_0(\epsilon)} +
\int_{-\infty}^{\infty} du\ \frac{K_0(\epsilon,u) -
K_1(\epsilon,u)}{\Gamma_{\text{tr}}(u)}\ ,
\label{eq:5}
\ee
where
\be
K_{0,1}(\epsilon,u) = -\NF\left[n\left(\frac{u-\epsilon}{T}\right) +
f\left(\frac{u}{T}\right)\right]\, \frac{{\bar
V}''_{0,1}(u-\epsilon)}{\Gamma_0(\epsilon)}\ .
\label{eq:6}
\ee
We will be interested in problems where the frequency in the spectrum $V''({\bf
p},u)$ scales as a power of the wave number, so that ${\bar V}''_1(u)/{\bar
V}''_0(u) \to 0$ for $u\to 0$. Since both the frequency $u$ and the energy
$\epsilon$ scale as the temperature, this implies that, effectively, $K_1/K_0
\to 0$ for $T\to 0$. This observation will be important later.

\section{Analysis of the integral equation}
\label{sec:3}

From Sec.\ \ref{sec:2} we see that we need to analyze the integral equation
\bse
\be
\varphi(\epsilon) = \frac{1}{\Gamma_0({\epsilon})} + \int_{-\infty}^{\infty} du
\left[K_0(\epsilon,u) - K_1(\epsilon,u)\right] \varphi(u)
\label{eq:7a}
\ee
where the first part of the kernel has the property
\be
\int_{-\infty}^{\infty} du\ K_0(\epsilon,u) = 1,
\label{eq:7b}
\ee
\ese
which follows from Eq.\ (\ref{eq:3}). Equation (\ref{eq:7a}) is a Fredholm
equation of the second kind, and in what follows we will use the known
properties of this class of integral equations \cite{Smirnov__5}. We will refer
to equations in Chapter 1 of Ref. \cite{Smirnov__5} as (S xx), and to theorems
by section number, (S Sec.n Theorem m).

\subsection{The homogeneous equation with kernel $K_0$}
\label{subsec:3.1}

We first consider the homogeneous integral equation
\be
\varphi(\epsilon) = \lambda \int_{-\infty}^{\infty} du\
K_0(\epsilon,u)\,\varphi(u).
\label{eq:8}
\ee
It follows from Eq.\ (\ref{eq:7b}) that $\lambda = \lambda_0 = 1$ is an
eigenvalue of the kernel $K_0$ with an eigenfunction $\varphi_0(\epsilon)
\equiv 1$, (S 34,35). Note that it also follows from Eq.\ (\ref{eq:7b}) that
$\int_{-\infty}^{\infty} d\epsilon\,du\, K_0(\epsilon,u) = \infty$. This means
that the sufficient condition (S 40ff) for an iterative solution of Eq.\
(\ref{eq:7a}) is badly violated, which suggests that iteration will not
converge. This in turn suggests that the behavior of $\Gamma_{\text{tr}}$ will
be qualitatively different from that of $\Gamma_0$.

It is also useful to consider the adjoint of Eq. (\ref{eq:8}),
\be
\psi(\epsilon) = \lambda \int_{-\infty}^{\infty} du\ K_0(u,\epsilon)\,\psi(u).
\label{eq:9}
\ee
From (S Sec.10 Theorem 9) it follows that for $\lambda=\lambda_0=1$ Eq.
(\ref{eq:9}) has a nonvanishing solution; i.e., there exists a nonvanishing
function $\psi_0(\epsilon)$ that obeys
\be
\psi_0(u) = \int_{-\infty}^{\infty} d\epsilon\ \psi_0(\epsilon)\
K_0(\epsilon,u).
\label{eq:10}
\ee

\subsection{The inhomogeneous equation with kernel $K_0$}
\label{subsec:3.2}

We next consider the inhomogeneous equation
\be
\varphi(\epsilon) = \frac{1}{\Gamma_0(\epsilon)} + \lambda
\int_{-\infty}^{\infty} du\ K_0(\epsilon,u)\,\varphi(u).
\label{eq:11}
\ee
(S Sec.10 Theorem 10) says that a necessary and sufficient condition for the
existence of a solution of Eq. (\ref{eq:11}) is
\be
\int_{-\infty}^{\infty} d\epsilon\ \frac{1}{\Gamma_0(\epsilon)}\ \psi(\epsilon)
= 0
\label{eq:12}
\ee
for every solution $\psi(\epsilon)$ of Eq.\ (\ref{eq:9}). An explicit check of
this condition would require to explicitly find the solutions of Eq.
(\ref{eq:9}), i.e., the left eigenfunctions of the kernel $K_0$, which is
difficult. However, we will see in what follows that one can construct
perturbative solutions of Eq. (\ref{eq:7a}) that do not exist as $K_1 \to 0$,
which implies that Eq. (\ref{eq:12}) cannot hold for all solutions of Eq.
(\ref{eq:9}).

\subsection{A perturbative eigenvalue of the full kernel}
\label{subsec:3.3}

Now we consider the following generalization of Eq. (\ref{eq:7a}),
\be
\varphi(\epsilon) = \frac{1}{\Gamma_0({\epsilon})} +
\lambda\int_{-\infty}^{\infty} du \left[K_0(\epsilon,u) - \alpha
K_1(\epsilon,u)\right] \varphi(u),
\label{eq:13}
\ee
which reduces to Eq. (\ref{eq:7a}) for $\lambda = \alpha = 1$. Our goal is to
use $\alpha$ as a small parameter. It follows from the observation at the end
of Sec. \ref{sec:2} that working to leading order in $\alpha$ and then setting
$\alpha=1$ will yield a result that is valid to leading order in $T$ as $T \to
0$.

We know from Sec. \ref{subsec:3.1} that for $\alpha=0$, $\lambda=\lambda_0=1$
is an eigenvalue of the kernel with eigenfunction $\varphi_0(\epsilon)\equiv
1$. For the generalization of the eigenproblem of Sec. \ref{subsec:3.1} to the
full kernel $K = K_0 - \alpha K_1$,
\be
\varphi_0(\epsilon) = \lambda_0 \int_{-\infty}^{\infty} du\ K(\epsilon,u)\
\varphi_0(u),
\label{eq:14}
\ee
this suggests a perturbative expansion
\bse
\label{eqs:15}
\be
\lambda_0 = \lambda_0^{(0)} + \alpha \lambda_0^{(1)} + O(\alpha^2)
\label{eq:15a}
\ee
with $\lambda_0^{(0)}=1$, and
\be
\varphi_0(\epsilon) = \varphi_0^{(0)}(\epsilon) + \alpha
\varphi_0^{(1)}(\epsilon) + O(\alpha^2),
\label{eq:15b}
\ee
\ese
with $\varphi_0^{(0)}(\epsilon) \equiv 1$. Similarly, for the solution of the
adjoint eigenproblem,
\bse
\label{eqs:16}
\be
\psi_0(u) = \lambda_0 \int_{-\infty}^{\infty} d\epsilon\ \psi_0(\epsilon)\
K(\epsilon,u),
\label{eq:16a}
\ee
we write
\be
\psi_0(\epsilon) = \psi_0^{(0)}(\epsilon) + \alpha \psi_0^{(1)}(\epsilon) +
O(\alpha^2),
\label{eq:16b}
\ee
\ese
where $\psi_0^{(0)}(\epsilon)$ is the function that solves Eq.\ (\ref{eq:10}).
Although we do not know this function explicitly, we have ascertained its
existence in Sec. \ref{subsec:3.1}.

Inserting these series into Eq. (\ref{eq:13}), multiplying with the function
$\psi_0(\epsilon)$, and comparing coefficients yields
\bse
\label{eqs:17}
\bea
\lambda_0 &=& 1 + \alpha \int_{-\infty}^{\infty} d\epsilon\
\psi_0^{(0)}(\epsilon)
[\Gamma_1(\epsilon)/\Gamma_0(\epsilon)]/\int_{-\infty}^{\infty} d\epsilon\
\psi_0^{(0)}(\epsilon)
\nonumber\\
&&+ O(\alpha^2).
\label{eq:17a}
\eea
Here we have defined
\be
\Gamma_1(\epsilon) = \Gamma_0(\epsilon) \int_{-\infty}^{\infty} du\
K_1(\epsilon,u).
\label{eq:17b}
\ee
We see that the first order correction to the eigenvalue $\lambda_0$ is given
by an average of the rate ratio $\Gamma_1/\Gamma_0$ with the zeroth order left
eigenfunction $\psi_0^{(0)}$ as the weight:
\be
\lambda_0 = 1 +
\alpha\,\left\langle\Gamma_1/\Gamma_0\right\rangle_{\psi_0^{(0)}},
\label{eq:17c}
\ee
or
\be
\lambda_0^{(1)} = \left\langle\Gamma_1/\Gamma_0\right\rangle_{\psi_0^{(0)}},
\label{eq:17d}
\ee
\ese
where $\langle f\rangle_{\psi} = \int_{-\infty}^{\infty} d\epsilon\
f(\epsilon)\,\psi(\epsilon) / \int_{-\infty}^{\infty} d\epsilon\
\psi(\epsilon)$.

\subsection{A perturbative solution of the full integral equation}
\label{subsec:3.4}

The desired solution of Eq. (\ref{eq:13}) can now be written in terms of a
resolvent $R(\epsilon,u)$, see (S 46),
\be
\varphi(\epsilon) = \frac{1}{\Gamma_0(\epsilon)} + \int_{-\infty}^{\infty} du\
R(\epsilon,u)\ \frac{1}{\Gamma_0(u)}\ ,
\label{eq:18}
\ee
and the resolvent can be expanded in a Laurent series (S Sec.8),
\be
R(\epsilon,u) = \frac{a_{-1}(\epsilon,u)}{1-\lambda_0} +
O\left((1-\lambda_0)^0\right).
\label{eq:19}
\ee
Since $1-\lambda_0 = O(\alpha)$, keeping only the leading term in the Laurent
series is consistent with working to leading order in $\alpha$. The residue
$a_{-1}$ satisfies the integral equation (S Sec.8)
\bea
a_{-1}(\epsilon,u) &=& \lambda_0 \int_{-\infty}^{\infty} dt\ K(\epsilon,t)\
a_{-1}(t,u) \nonumber\\
&=& \int_{-\infty}^{\infty} dt\ K_0(\epsilon,t)\ a_{-1}(t,u) + O(\alpha).
\label{eq:20}
\eea
If we define
\bse
\label{eqs:21}
\be
A_{-1}(\epsilon) = \int_{-\infty}^{\infty} du\ a_{-1}(\epsilon,u)/\Gamma_0(u)
\label{eq:21a}
\ee
we have
\be
A_{-1}(\epsilon) = \int_{-\infty}^{\infty} dt\ K_0(\epsilon,t)\ A_{-1}(t) +
O(\alpha).
\label{eq:21b}
\ee
To lowest order in $\alpha$, this is just Eq. (\ref{eq:8}) with $\lambda=1$,
which has a solution
\be
A_{-1}(\epsilon) = {\text{const.}} + O(\alpha) \equiv A_{-1} + O(\alpha).
\label{eq:21c}
\ee
\ese
The solution of the integral equation can now be written
\be
\varphi(\epsilon) = \frac{A_{-1}}{\alpha}\ \frac{1}{\lambda_0^{(1)}} +
O(\alpha^0).
\label{eq:22}
\ee
We see that to leading order in the Laurent expansion in $\alpha$, $\varphi$
does not depend on $\epsilon$. We also see that the solution does not exist for
$\alpha=0$, as we had anticipated in Sec.\ \ref{subsec:3.2} above.

The remaining task is to determine the constant $A_{-1}$, which is not fixed by
the integral equation (\ref{eq:21b}). Let us consider the functions
$\varphi(\epsilon)$ and $1/\Gamma_0(\epsilon)$ as elements
$\vert\varphi\rangle$ and $\vert 1/\Gamma_0\rangle$, respectively, of a linear
space, and the kernel $K$ a linear operator in that space. The integral
equation (\ref{eq:13}) then reads
\be
\vert\varphi\rangle = \vert 1/\Gamma_0\rangle + K\vert\varphi\rangle.
\label{eq:23}
\ee
Multiplying from the left with the left eigenfunction $\langle\psi_0\vert$ of
$K$, we have
\bea
\hskip 20pt\langle\psi_0\vert\varphi\rangle &=& \langle\psi_0\vert
1/\Gamma_0\rangle +
\langle\psi_0\vert K\vert\varphi\rangle \nonumber\\
&=& \langle\psi_0\vert 1/\Gamma_0\rangle + \lambda_0
\langle\psi_0\vert\varphi\rangle.
\label{eq:24}
\eea
From Eq. (\ref{eq:22}) we know that, to lowest order in $\alpha$,
$\vert\varphi\rangle$ is a multiple of $\vert\varphi_0^{(0)}\rangle$,
\be
\vert\varphi\rangle = \frac{1}{\alpha}\ \frac{A_{-1}}{\lambda_0^{(1)}}\
\vert\varphi_0^{(0)}\rangle + O(\alpha^0).
\label{eq:25}
\ee
To leading order in $\alpha$ we thus have
\be
(1-\lambda_0)\ \langle\psi_0^{(0)}\vert\phi_0^{(0)}\rangle\ [1+O(\alpha)] =
\langle\psi_0^{(0)}\vert 1/\Gamma_0\rangle,
\label{eq:26}
\ee
which yields
\be
A_{-1} = \frac{-\langle\psi_0^{(0)}\vert
1/\Gamma_0\rangle}{\langle\psi_0^{(0)}\vert \varphi_0^{(0)}\rangle}.
\label{eq:27}
\ee
Using the notation defined in Eqs. (\ref{eqs:17}) we have
\be
\varphi(\epsilon) = \frac{1}{\alpha}\ \frac{\langle
1/\Gamma_0\rangle_{\psi_0^{(0)}}}{\langle\Gamma_1/\Gamma_0\rangle_{\psi_0^{(0)}}}
+ O(\alpha^0).
\label{eq:28}
\ee
Finally, using the observation made after Eq. (\ref{eq:13}) we can put
$\alpha=1$ and write
\be
\varphi(\epsilon) = \frac{\langle
1/\Gamma_0\rangle_{\psi_0^{(0)}}}{\langle\Gamma_1/\Gamma_0\rangle_{\psi_0^{(0)}}}
+ O(\Gamma_1^0).
\label{eq:29}
\ee
This is the desired solution of the integral equation (\ref{eq:7a}) to leading
order in the function $\Gamma_1(\epsilon)$ defined in Eq. (\ref{eq:17b}). The
only assumption we have made is that the left eigenfunction
$\psi_0^{(0)}(\epsilon)$, which we have not explicitly determined, falls off
sufficiently fast for large $\epsilon$ to ensure the existence of the averages
in Eq. (\ref{eq:27}).

The above result is exact. Let us compare it with the popular approximation
\cite{Mahan_1981__3} that replaces $\varphi(u)$ under the integral in Eq.
(\ref{eq:7a}) by $\varphi(\epsilon)$. This turns the integral equation into an
algebraic one, and the solution is
\be
\varphi(\epsilon) \approx 1/\Gamma_1(\epsilon).
\label{eq:30}
\ee
This has the same structure in terms of $\Gamma_1$ as the exact solution
(\ref{eq:28}), and leads to the same temperature dependence of the
conductivity, as we will see. However, it is uncontrolled and cannot be
justified other than by a comparison with the exact solution provided above, or
with a solution of the Boltzmann equation.

\section{The electrical conductivity}
\label{sec:4}

\subsection{The Drude formula}
\label{subsec:4.1}

We now can write the conductivity as follows. From Eq. (\ref{eq:1}) we have
\be
\sigma = \frac{\ne e^2}{2\me} \int_{-\infty}^{\infty} \frac{d\epsilon}{4T}\
\frac{1}{\cosh^2(\epsilon/2T)}\ \frac{1}{\Gamma_{\text{tr}}(\epsilon)}\ .
\label{eq:31}
\ee
Equation (\ref{eq:29}) shows that, to lowest order in $\Gamma_1$,
$\Gamma_{\text{tr}}$ is independent of $\epsilon$ and given by
\be
\Gamma_{\text{tr}} =
\frac{\langle\Gamma_1/\Gamma_0\rangle_{\psi_0^{(0)}}}{\langle
1/\Gamma_0\rangle_{\psi_0^{(0)}}} + O(\Gamma_1^2).
\label{eq:32}
\ee
We thus have as our final result the Drude formula
\bse
\label{eqs:33}
\be
\sigma = \frac{\ne e^2\tau_{\text{tr}}}{\me},
\label{eq:33a}
\ee
with a transport relaxation rate
\be
\frac{1}{\tau_{\text{tr}}} = 2\Gamma_{\text{tr}}\ .
\label{eq:33b}
\ee
Here $\Gamma_{\text{tr}}$ is given by Eq. (\ref{eq:32}). It depends on
$\Gamma_0$ from Eq. (\ref{eq:3}), $\Gamma_1$ from Eq. (\ref{eq:17b}), the
kernels $K_0$ and $K_1$ defined by Eq. (\ref{eq:6}), and the left eigenfunction
$\psi_0^{(0)}$ of $K_0$ that is defined as the solution of the equation
\be
\psi_0^{(0)}(u) = \int_{-\infty}^{\infty} d\epsilon\ \psi_0^{(0)}(\epsilon)\
K_0(\epsilon,u).
\label{eq:33c}
\ee
\ese

\subsection{Two physical examples}
\label{sec:4.2}

We finally illustrate our result with two examples. One is the well-known
result for the conductivity due to electron-phonon scattering, the other, the
conductivity due to the scattering of helical magnon excitations in
helimagnets. In both cases, the effective potentials ${\bar V}''_{0,1}$ that
determine the kernels have the following power-law form in the limit of low
frequencies:
\be
{\bar V}''_m(u) = \frac{-C_m}{\NF}\ \vert u\vert^{n_m-1}\ \sgn u, \quad
(m=0,1),
\label{eq:34}
\ee
where $C_0$ and $C_1$ are positive constants. For acoustic phonons, the
exponents $n_m$ are \cite{AGD}
\bse
\label{eqs:35}
\be
n_0 = 3, \quad n_1 = 5 \quad (\text{phonons}).
\label{eq:35a}
\ee
For helimagnons, the corresponding values are \cite{BKR__4}
\be
n_0 = 3/2, \quad n_1 = 5/2 \quad (\text{helimagnons}).
\label{eq:35b}
\ee
\ese
For the relaxation rates $\Gamma_{0,1}$ this yields
\bse
\label{eqs:36}
\be
\Gamma_m(\epsilon) = T^{n_m}\ \gamma_m(\epsilon/T), \quad (m=0,1),
\label{eq:36a}
\ee
where
\bea
\gamma_m(x) &=& C_m \int_{0}^{\infty} du\ u^{n_m-1}\ \left[2n(u) + f(u+x)
\right. \nonumber\\
&& \hskip 100pt \left. + f(u-x)\right].
\label{eq:36b}
\eea
\ese
Finally, it follows from Eqs. (\ref{eq:6}) and (\ref{eq:33c}) that
$\psi_0^{(0)}(\epsilon)$ is a function of $\epsilon/T$ only. For the transport
relaxation time, and hence for the conductivity, we thus have the a temperature
dependence from electron-phonon scattering:
\bse
\label{eqs:37}
\be
\sigma \propto \tau_{\text{tr}} \propto 1/T^5 \quad (\text{phonons}),
\label{eq:37a}
\ee
This is the well-known Bloch-Gr{\"u}neisen law \cite{Wilson_1954__1}. The same
power law is obtained by means of the uncontrolled approximation (\ref{eq:30})
\cite{Mahan_1981__3}, but the prefactor is different. For scattering from
helimagnons we find
\be
\sigma \propto \tau_{\text{tr}} \propto 1/T^{5/2} \quad (\text{helimagnons}).
\label{eq:37b}
\ee
\ese
Again, the uncontrolled approximation yields the same $1/T^{5/2}$ temperature
behavior \cite{BKR__4}. The above considerations ascertain that this is indeed
the exact low-temperature contribution to the electrical conductivity due to
electron-helimagnon scattering.

\section{Conclusion}

In conclusion, we have considered the Kubo formula for the electrical
conductivity due to the scattering of electrons by bosonic excitations. We have
given an exact solution of the integral equation for the vertex function that
determines the transport relaxation rate. This exact solution has the
qualitatively same behavior as the result of a popular, but uncontrolled,
approximation. For the scattering of electrons by acoustic phonons this yields
the usual Bloch-Gr{\"u}neisen law for the electrical resistivity, $\rho \propto
T^5$, and for the helimagnon contribution to the conductivity in helical
magnets it confirms the result $\rho \propto T^{5/2}$ that had been previously
obtained by means of solving the integral equation approximately. All of these
results are consistent with what is known from solutions of the Boltzmann
equation.

\section*{Acknowledgments}

We thank N. Christopher Phillips for a helpful discussion. This work was
supported by the NSF under grant Nos. DMR-05-30314 and DMR-05-29966.

\end{document}